\documentclass[prd, preprint, nofootinbib]{revtex4}
\usepackage{graphicx}
\usepackage{amsfonts}
\usepackage{latexsym}
\usepackage{amssymb}
\usepackage{epsfig}

\begin{document}

\title{Entropy production by domain wall decay in the NMSSM}

\author{Hironori Hattori, Tatsuo Kobayashi, Naoya Omoto}
 \affiliation{Department of Physics, Hokkaido University, Sapporo 060-0810, Japan}

\author{Osamu Seto}
 \affiliation{Department of Life Science and Technology,
  Hokkai-Gakuen University, Sapporo 062-8605, Japan}

%
\begin{abstract}
We consider domain walls in the $Z_3$ symmetric NMSSM.
The spontaneous $Z_3$ discrete symmetry breaking produces domain walls, 
and the stable domain walls are problematic.
Thus, we assume the $Z_3$ symmetry is slightly but explicitly broken
 and the domain walls decay.
Such a decay causes a large late-time entropy production.
We study its cosmological implications on unwanted relics such as 
moduli, gravitino, LSP and axion.
\end{abstract}

\preprint{HGU-CAP-039} 
\preprint{EPHOU-15-014}

\vspace*{3cm}
\maketitle


\section{Introduction}

Supersymmetric extension of the standard model (SM) is one of 
candidates for TeV-scale physics, 
because supersymmetry (SUSY) can stabilize a large hierarchy.
The minimal supersymmetric standard model (MSSM) is quite interesting 
because of its minimality, 
and various phenomenological aspects have been studied.
However, from theoretical point of view, it has a problem.
The MSSM includes supersymmetric mass terms of Higgs superfields, $H_u$ and $H_d$
i.e. the so-called $\mu$-term, $\mu H_u H_d$, in the superpotential.
It must be comparable with soft SUSY breaking masses 
in order to realize successfully the electroweak symmetry breaking.
However, the $\mu$-term and soft SUSY breaking terms, 
in general, have origins different from each other.
Why are these comparable with each other ?
That is the so-called $\mu$-problem~\cite{Kim:1983dt}.

The next-to-minimal supersymmetric standard model (NMSSM) is 
an extension of the MSSM by adding a singlet superfield $S$~\cite{Fayet:1974pd}
(see for a review Ref.~\cite{Ellwanger:2009dp}).
Then, the NMSSM superpotential has $\lambda S H_u H_d$.
Also, we impose the $Z_3$ symmetry, which forbids 
dimensionful parameters in the superpotential.
Dimensionful parameters appear only as soft SUSY breaking parameters.
Thus, vacuum expectation values (VEVs) of Higgs and singlet fields 
are determined by soft SUSY breaking terms.
That is, the $\mu$-problem is solved, and 
the effective $\mu$-term is generated as $\mu= \lambda \langle S \rangle$.

In the NMSSM, the Higgs sector as well as the neutralino sector 
has a richer structure than one in the MSSM, because of inclusion 
of the singlet superfield $S$.
Also, the NMSSM can raise the SM-like Higgs boson mass.
At any rate, heavier superpartner masses such as ${\cal O}(1)-{\cal O}(10)$ TeV 
may be favorable.
We may need fine-tuning to realize a little hierarchy between 
the electorweak scale and SUSY breaking scale.
However, such a fine-tuning can be improved in a certain mediation mechanism, 
e.g. in the TeV-scale mirage mediation scenario \cite{Kobayashi:2012ee}.\footnote{
See for phenomenological aspects of MSSM in the TeV-scale mirage mediation scenario \cite{Choi:2005hd} 
and for generic mirage mediation \cite{Choi:2004sx,Choi:2005uz,Endo:2005uy}.}

The $Z_3$ symmetry is important to forbid dimensionful parameters in the superpotential 
and to solve the $\mu$-problem.
However, it is problematic.
VEVs of the Higgs scalar and singlet break spontaneously the $Z_3$ symmetry.
In general, when a discrete symmetry is spontaneously broken, 
domain walls appear.
They would dominate the energy density of the Universe and change the standard cosmology drastically. 
Thus, the exact $Z_3$ symmetry and the stable domain walls are problematic \cite{Zeldovich:1974uw}.
See for the NMSSM \cite{Abel:1995wk} .

Here, we assume that the $Z_3$ symmetry is broken explicitly, 
 but its breaking size is much smaller than the electroweak scale.
Then, the domain walls are unstable.
They may dominate the energy density of the Universe at a certain period 
but decay.
It has important effects on thermal history. (See e.g. Ref~\cite{Kadota:2015dza}.)
In this paper, we study implications of unstable domain walls in the NMSSM.
In general, SUSY models have other problems due to moduli, gravitino
 and the lightest superparticle (LSP).
For example, in the gravity mediation scenario,
 moduli and gravitino masses 
would be comparable with masses of superpartners in the visible sector.
When those are of ${\cal O}(1)-{\cal O}(10)$ TeV, they affect 
successful big bang nucleosynthesis (BBN), that is, the so-called 
moduli-problem and gravitino problem.
They could be  diluted  by decay of domain walls~\cite{Kawasaki:2004rx}.
Furthermore, even if the moduli and gravitino are heavier than superpartners
 in the visible sector, that would lead to another problem.
Indeed, in the mirage mediation mechanism~\cite{Choi:2004sx}, 
the gravitino is heavier by ${\cal O}(8\pi^2)$ than superpartners in the visible sector, 
and the modulus is also heavier by ${\cal O}(8\pi^2)$ than the gravitino.
In such a case, the moduli decay into the gravitino with a large rate and 
the gravitino decays into the LSP.
This overproduces non-thermally the LSP~\cite{Endo:2006zj}.
We need to dilute the moduli, gravitino and the LSP.
Also, in some other scenarios, the LSP such Bino-like neutralino has a large thermal relic density~.
The decay of domain walls, which was mentioned above, can produce a large entropy 
and dilute moduli and dark matter candidates 
in the NMSSM.

This paper is organized as follows.
In section 2, we study the domain wall solution in the NMSSM.
In section 3, we study cosmological evolution of unstable domain walls.
In sections 4 and 5, we study implications of the domain wall decay in two scenarios.
Section 6 is devoted to conclusion and discussion.

\section{Domain wall solution in the NMSSM}

\subsection{Domain wall solution in the $Z_3$ symmetric NMSSM}

We briefly review a domain wall solution of the Higgs potential in the $Z_3$ symmetric NMSSM
\footnote{
The full scalar potential includes superpartners of quarks and leptons, and 
it has several unrealistic vacua. 
We assume that taken SUSY breaking parameters in the full potential
 satisfy the condition to avoid such unrealistic vacua.
 (See e.g., Ref~\cite{Kanehata:2011ei}  and references therein.)}.
We adopt the convention for $H_u$, $H_d$ and $S$
 that the superfield and its lowest component are written by the same letter.
The superpotential terms including only $H_u$, $H_d$ and $S$ are written as
\begin{eqnarray}
W_{\rm Higgs}=\lambda S H_u H_d+\frac{\kappa}{3}S^3,
\end{eqnarray}
 where the $Z_3$ symmetry is imposed as mentioned.
The scalar potential is written by 
\begin{eqnarray}
V_{\rm Higgs}=\sum_{\phi_i=H_u, H_d, S} \left|\frac{\partial W}{\partial \phi_i} \right|^2 + V_D + V_{{\rm soft}},
\end{eqnarray}
where $V_D$ is the D-term potential due to $SU(2)\times U(1)_Y$ and 
$V_{\rm soft}$ denotes the soft SUSY breaking terms,
\begin{eqnarray}
 V_{{\rm soft}} = m_{H_u}^2 |H_u|^2+ m_{H_d}^2 |H_u|^2 
 + \frac13 \kappa A_{\kappa}S^3 + \lambda A_{\lambda} H_u H_dS+ h.c. 
\end{eqnarray}
Only the neutral components develop their VEVs, and their scalar potential is written explicitly by 
\begin{eqnarray}
V_{\rm Higgs} &=& \left|\kappa S^2 - \lambda  H_u^0H_d^0 \right|^2+m_{H_u}^2 |H_u^0|^2+m_{H_d}^2 |H_d^0|^2+m_S^2 \left|S \right|^2  
 +\left| \lambda \right|^2\left|S \right|^2(|H_d^0|^2+|H_u^0|^2) \nonumber \\
& &  +\frac{ g^2 +g'^2 }{8} \left( |H_u^0|^2-|H_d^0|^2 \right) ^2 
 +\frac13 \kappa A_{\kappa}S^3-\lambda A_{\lambda} H_u^0 H_d^0 S+ h.c.,
\end{eqnarray}
 where $g$ and $g'$ are the $SU(2) $ and $U(1)_Y$ gauge couplings, respectively.
Here, we assume that all of  $\lambda$, $\kappa$, $A_\lambda$ and $A_\kappa$ are real.

The potential minima are obtained by analyzing the stationary conditions, 
\begin{eqnarray}
\frac{\partial V_{\rm Higgs}}{\partial H_u^0} = \frac{\partial V_{\rm Higgs}}{\partial H_d^0}
= \frac{\partial V_{\rm Higgs}}{\partial S} =0,
\end{eqnarray}
and these VEVs lead to the successful electroweak symmetry breaking, 
where the effective $\mu$ term is obtained as $\mu = \lambda \langle S \rangle$.
Since the scalar potential has the $Z_3$ symmetry, three vacua are degenerate,
\begin{eqnarray}
\left(\left<S\right>,\left<H_u^0\right>,\left<H_d^0\right> \right) 
= \left(v_s e^{2\pi i m/3} ,v_{u} e^{2\pi i m/3} ,v_{d} e^{2\pi i m/3} \right),
\end{eqnarray}
with $m=0,1,2$, where all $v_s, v_u$ and $v_d$ are real with
 $v = \sqrt{v_u^2+v_d^2}\simeq 174$ GeV.
One of three degenerate vacua is selected in the vacuum, and then the $Z_3$ symmetry is broken spontaneously.
Then, the domain walls are generated.

First, we study the domain wall solution~\cite{Vilenkin:2000jqa}.
We fix field values of radial directions of $S, H_u$ and $H_d$, and
 discuss a field equation for the phase degree of freedom $\phi$, 
\begin{eqnarray}
\left(S,H_u^0,H_d^0\right) 
= \left(v_s e^{{\mathrm i}\phi} ,v_{u} e^{{\mathrm i}\phi} ,v_{d} e^{{\mathrm i}\phi} \right).
\end{eqnarray}
The potential of $\phi$ can be obtained from $V_{\rm Higgs}$ as
\begin{eqnarray}
V_{}(\phi)&=& -2\left(\frac13 \left| \kappa A_{\kappa}v_s^3\right| +\lambda A_{\lambda} v_sv_uv_d\right)\cos (3\phi) 
+ V_0,
\end{eqnarray}
where $V_0$ denotes $\phi$-independent terms.
The first term would be dominant when $A_\kappa \sim A_\lambda$, $\lambda \sim \kappa$ and 
$v_s^2 \gg v_u v_d$.
Also, the kinetic term of $\phi$ is written by 
\begin{eqnarray}
\mathcal{L}_{{\rm kinetic}}(\phi)
&=&\eta^2(\partial_{\mu} \phi)(\partial^{\mu} \phi),
\end{eqnarray}
 with $\eta^2 = v_s^2+v_u^2+v_d^2$.

For simplicity, we consider a planar domain wall orthogonal to the $z$-axis, 
$\phi(z)$.
Then, the field equation,
\begin{eqnarray}
\partial_{\mu}\frac{\partial \mathcal{L}_{{\rm kinetic}}}{\partial_{\mu}(\partial \phi)}+\frac{\partial V_{{\rm VEV}}}{\partial \phi}=0,
\end{eqnarray}
 can be written by 
\begin{eqnarray}
\frac{ \mathrm{d}^2 \phi }{ \mathrm{d}z^2 }-\frac{1}{3B^2} \sin (3\phi)&=&0,
\end{eqnarray}
 with
\begin{eqnarray}
\left( \frac{1}{B}\right)^2 = \frac{9\left(|\frac13 \kappa A_{\kappa}v_s^3|+\lambda A_{\lambda} v_sv_uv_d\right)}{\eta^2}.
\label{def:B}
\end{eqnarray}
The first term in the numerator of the left hand side of Eq.~(\ref{def:B})
 is dominant when $v_s^2 \gg v_uv_d$.
We set the boundary condition such that $\phi = 2 \pi n/3$ at $z \rightarrow - \infty$ and 
$\phi = 2 \pi (n+1)/3$ at $z \rightarrow + \infty$ with $n=0,1,2$.
By solving the above field equation with this boundary condition, the domain wall solution is derived as
\begin{eqnarray}
\phi &=& \frac{2 n \pi}{3}+ \frac43 \arctan\left(\mathrm{e}^{\pm \frac1B(z-z_0)}\right),
\end{eqnarray}
where $B$ corresponds to the width of the domain wall.
Figure 1 shows this solution for $n=0$.

Now, we can estimate the domain wall tension
\begin{eqnarray}
\sigma &=& \int dz \rho_{{\rm wall}}(z)= \int dz \left( \left|\frac{ \mathrm{d}S }{ \mathrm{d}z } \right|^2+ \left|\frac{ \nonumber \mathrm{d}H_u^0 }{ \mathrm{d}z } \right|^2+ \left|\frac{ \mathrm{d}H_d^0 }{ \mathrm{d}z } \right|^2+V(\phi)\right) \\ \nonumber
&=&\frac{16}{9}\frac{\eta^2}{B} .
\end{eqnarray}
Thus, we can estimate 
\begin{eqnarray}
 \sigma \simeq \frac{16}{3\sqrt{3}}v_s^2\sqrt{\kappa A_\kappa v_s} = 
\frac{16}{3\sqrt{3}}\frac{\mu^2}{\lambda^2}\sqrt{\frac{\kappa}{\lambda} A_\kappa \mu},
\end{eqnarray}
 for $v_s^2 \gg v_uv_d$.
The size of $\mu$ is of the SUSY breaking
 scale.~\footnote{When $\mu$ is much larger than the elwctroweak scale,
 we have the fine-tuning problem to derive the Z-boson mass $m_Z$
 from $m_{H_u}^2, \mu,$ and $m_{H_d}^2$.
However, in a certain mediation such as the TeV-scale mirage mediation contributions
 due to $\mu$ and $m_{H_d^2}$ cancel each other in $m_Z$,
 and $m_Z$ is independent of $\mu$.
Without severe fine-tuning $\mu$ can be larger than the electroweak
 scale, e.g. $\mu = {\cal O}(1) $TeV~\cite{Kobayashi:2012ee}.}
The couplings $\lambda$ and $\kappa$ must be of ${\cal O}(0.1)$ or less
 at the electorweak scale such that they do not blow up below a high energy scale
 such as the GUT or Planck scale.
Thus, the size of $\sigma^{1/3}$ would be of the SUSY breaking scale or larger.
Figure 2 shows an example of $\rho_{DW}(z)$.\\

%
\begin{figure}[ht]
\begin{center}
\epsfig{figure=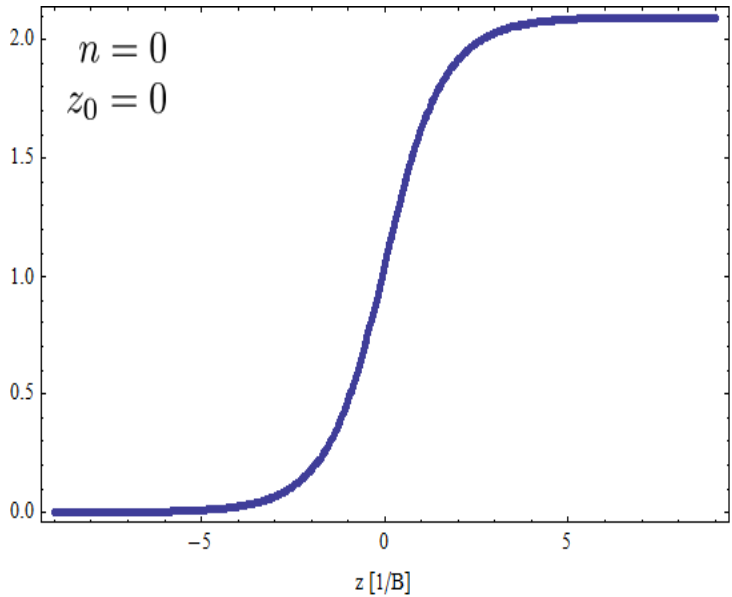, width=10cm,height=8cm,angle=0}
\end{center}
\caption{The phase of scaler field$(S(z), H_u(z), H_d(z))$ of planer domain wall solution. 
Here we take $n=0$, $z_0=0$, and normalize z-axis by $1/B$ (Eq.~(\ref{def:B})).
 }
\label{Fig:phase}
\end{figure}
%

%
\begin{figure}[ht]
\begin{center}
\epsfig{figure=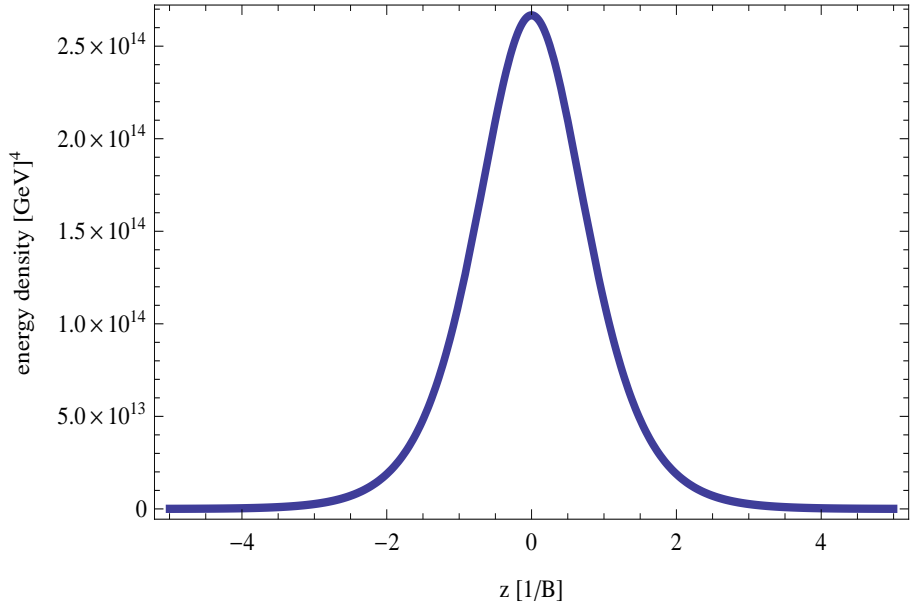, width=12cm,height=8cm,angle=0}
\end{center}
\caption{Spatial configuration of a domain wall energy density for $\lambda=\kappa=0.01$, 
$A_{\lambda}=A_{\kappa}=10~{\rm TeV}$, $\mu=1~{\rm TeV}$, $\tan \beta =10$. The z-axis is normalized by $1/B$.
 }
\label{Fig:energydensity}
\end{figure}
%

\subsection{Decaying domain wall by $Z_3$ breaking}

Formed domain walls are stretched by the cosmic expansion and
 smoothed by interactions with particles in the background thermal plasma.
The energy density of domain walls $\rho_{DW}$ and its pressure $p_{DW}$ can be read
 from the averaged energy momentum tensor of domain walls.
The equation of state of domain walls is given by
\begin{eqnarray}
  p_{DW} = \left(v^2-\frac{2}{3}\right)\rho_{DW} ,
\end{eqnarray}
 with $v$ being the averaged velocity of walls~\cite{KolbTurner}.
The dynamics depends on $v$.
In one extremal limit, non-relativistic limit or static limit with $v=0$,
 the energy density behaves 
\begin{eqnarray}
\rho_{DW} \propto a^{-1},
\end{eqnarray}
 where $a(t)$ is the scale factor of the Universe.
Such domain wall network is sometime referred to as ``frustrated domain wall''.
Such a frustrated domain wall dominated Universe causes acceralating expansion 
 because of $w = p/\rho = -2/3 < -1/3$.
On the other hand, for $v^2 \geq 1/3$ where $w \geq -1/3$ is realized,
 the cosmic expansion is not acceralating.

In fact, the dynamics of domain walls has been investiagted and
 many detailed numerical simulations show that
 the dynamics of domain wall network is relaxed at a late time to so-called scaling regime,
 where the typical length scale $\xi$ of the system stays of the Hubble
 radius $H^{-1}$~\cite{Press:1989yh,Hindmarsh:2002bq,Garagounis:2002kt,
Leite:2011sc,Leite:2012vn,Hiramatsu:2013qaa}.
Then, the energy density of domain walls also scales as~\cite{Hiramatsu:2013qaa}
\begin{eqnarray}
  \rho_{DW} \simeq \frac{\sigma}{t}.
\end{eqnarray}

The energy density of domain wall decreases slower than any other ``matter'' or radiation
 in the scaling solution~\footnote{In the static limit $v=0$, it is furtehr slower.}.
Thus, at some point, the energy density of domain walls dominates that in the Universe. 
This is the domain wall problem~\cite{Zeldovich:1974uw}.

Thus, the stable domain wall in the $Z_3$ symmetric NMSSM is problematic~\cite{Abel:1995wk}.
In this paper, we consider a tiny but explicit breaking of the $Z_3$ discrete symmetry 
 so that domain walls might have a long life time but finally decay.
In fact, the decay of domain walls after domain wall domination
 has an interesting cosmological implication,
 namely the dilution of unwanted relics
 by late time entropy production~\cite{Kawasaki:2004rx}.
 
Few numerical detailed study on dymanics of the domain walls network
 in a domain wall dominated Universe has been done.
Hence, the domain wall dynamics in a domain wall dominated Universe after
 its scaling behavior is uncertain.
One likely possibility is that the scale of the system remains
 of the order of the Hubble radius as in the scaling regime
 after domain wall domination too.
This can be realized for the equation of the state $w\simeq -1/3$.
Thus, in the most of the following analysis, we assume this.
On the other hand, there is another possibility that 
 the dynamics after the domination would be frozen as suggested in Ref.~\cite{Leite:2011sc},
 where $\xi \propto a(t)$ and $\rho \propto a(t)^{-1}$ are realized
 as in the non-relativistic limit.
We briefly discuss results for this latter case too.

Before closing this subsection,
 here we briefly note some examples of the $Z_3$ symmetry breaking
 in the literature for information.
In Ref.~\cite{Panagiotakopoulos:1998yw}, Panagiotakopoulos and Tamvakis
 proposed adding extra symmetries which consistently allows
 to induce a tiny enough tad pole term
\begin{equation}
  \Delta V \sim \frac{1}{(16\pi^2)^n} m_{SUSY}^3 (S+S^*),
\end{equation}
 where $m_{SUSY}$ is a soft SUSY breaking mass and $n$ is a power of loop inducing this term,
 in the scalar potential and the degeneracy of vacua is resolved.
Hamaguchi et al proposed another solution by intorducing hidden QCD theory,
 where the $Z_3$ symmetry becomes anomalous and is broekn
 by quantum effects~\cite{Hamaguchi:2011nm}.
In such a minor extension of $Z_3$ symmetric NMSSM,
 the domain walls become unstable.
Since the size of the $Z_3$ breaking term is highly model dependent
 and the main purpose of  this paper is
 to study cosmological effects of late time domain walls decay,
 the decay rate of a domain wall $\Gamma_{DW}$,
 which also parameterise the size of the $Z_3$ symmetry breaking,
 is treated as a free parameter.
Throughout this paper, in order to connect successful BBN,
 we take the domain wall decay temperature $T_d$ of a few MeV.
We note that the lower bound of the rehearting temperature
 by late decay objects is about a few MeV~\cite{Kawasaki:2000en,Hannestad:2004px,Ichikawa:2005vw}. 

\section{Cosmological evolution of unstable domain walls}

When doublet and/or singlet Higgs fields develop the VEVs,
 domain walls are formed.
%
As mentioned above, after certain dynamics,
 the domain wall network would be relaxed to be in the scaling solution.
In the scaling regime, the energy density of domain walls is given by
\begin{eqnarray}
\rho_{DW} \simeq \sigma H.
\label{eq:ehoDW:inscaling}
\end{eqnarray}
%


\subsection{Matter-dominated era to domain wall dominated era}

The first case we consider is that,
 at the domain wall formation time $H_i^{-1}$, 
 the Universe is dominated by the energy density of a matter $\rho_M$ such as
 a long-lived coherent oscillating moduli field.
In the scaling solution of domain wall,
 the energy density of domain walls relative to that of the background
 increases and eventually dominates the Universe. 
The domain wall energy density becomes equal to 
 one of the matter at $H_{eq}^{-1}$, which is estimated with Eq.~(\ref{eq:ehoDW:inscaling}) as 
\begin{eqnarray}
H_{eq} \simeq  \frac{\sigma}{3 M_P^2 } ,
\label{eq:H-eq:Scaling:MD}
\end{eqnarray}
where $M_P$ is the reduced Planck mass.
The condition that domain walls indeed dominate the Universe before those decay is expressed as
\begin{eqnarray}
H_{eq} > \Gamma_{DW}.
\end{eqnarray}
After $H_{eq}$, the domain walls dominate the energy density.

At the domain wall decay time $\Gamma_{DW}^{-1}$,
 the ratio of these energy densities is estimated as 
\begin{eqnarray}
 \left. \frac{\rho_M}{\rho_{DW}}\right|_{\Gamma_{DW}} = \left( \frac{\Gamma_{DW}}{H_{eq}}\right),
\end{eqnarray}
 from $a \propto t $, where we assume $\rho_{DW} \propto a^{-2}$
 during the domain wall domination between $H_{eq}$ and $\Gamma_{DW}$.
After the domain walls decay, the energy density of the matter is diluted as
\begin{eqnarray}
\frac{\rho_{M}}{s} = \frac{ 3 T_d}{4}\left( \frac{\Gamma_{DW}}{H_{eq}}\right)
 \simeq \frac{ 3 T_d}{4}\left( \frac{\pi^2 g_*(T_d) T_d^4}{10} \frac{M_P^2}{\sigma^2} \right)^{1/2} ,
\label{eq:rho/s:Scaling2}
\end{eqnarray}
 for the case that the domain wall decays earlier than the matter does.
Here, $g_*$ is the number of relativistic degrees of freedom.

\subsection{Radiation-dominated era to domain wall dominated era}

Next, we discuss the case
 that domain walls are formed in radiation-dominated Universe.
Both energy densities become comparable with each other at 
\begin{eqnarray}
H_{eq} \simeq  \frac{\sigma}{3 M_P^2},
\label{eq:H-eq:Scaling:RD}
\end{eqnarray}
 since domain walls are in the scaling solution.
%
The entropy density ratio of after- to before-domain wall decay is given by 
\begin{eqnarray}
 \Delta = \frac{s_{after}}{s_{before}} \simeq \frac{T_{eq}}{T_d}\left( \frac{H_{eq}}{\Gamma_{DW}}\right)
 \simeq \left(\frac{10 \sigma^2}{\pi^2 g_*(T_d) T_d^4 M_P^2} \right)^{3/4}
 \left(\frac{g_*(T_d)}{g_*(T_{eq})}\right)^{1/4},
\label{rad:delta1-2:Scaling2}
\end{eqnarray}
 for $\Delta \gg 1$.
We can obtain an entropy production 
\begin{eqnarray}
 \Delta \simeq 10 \left( \frac{\sigma^{1/3}}{ 50 \,{\rm TeV}}\right)^{9/2}
\left( \frac{2\,{\rm MeV}}{T_d}\right)^3 .
\label{rad:delta2:Scaling2}
\end{eqnarray}

One might think that the tension of about $100$ TeV looks somewhat too large.
However, for instance,
 in the MSSM-like region of the NMSSM with $\lambda\sim \kappa \ll 1$ and $v_s \gg v$,
 the domain wall tension
\begin{eqnarray}
 \sigma \simeq \frac{16}{3}\sqrt{\frac{2}{3}} \kappa v_s^3 ,
\end{eqnarray}
 can be of such an order with $\lambda\sim \kappa \sim 10^{-2}$
 and $v_s \sim 100$ TeV. 
Those results in the effective $\mu$ term and the singlino mass of about $1$ TeV.
Figure 3 shows the entropy density ratio of after- to before- domain wall decay for 
$\lambda=\kappa=0.01$, $T_d=3~{\rm MeV}$. 
The ratio increases as $\mu$ and $A_{\kappa}$ increase.

Such a large late-time entropy production can dilute unwanted relics
 such as gravitino, overproduced LSP as well as axion.\\
%
\begin{figure}[ht]
\begin{center}
\epsfig{figure=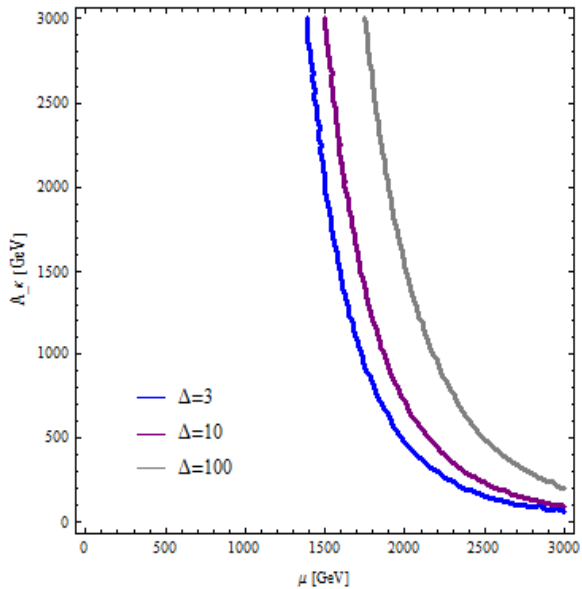, width=8cm,height=8cm,angle=0}
\end{center}
\caption{
The entropy density ratio $\Delta$ of after- to before- domain wall decay 
in radiation-dominated era to domain wall dominated era 
for $\lambda=\kappa=0.01$, $T_d=3~{\rm MeV}$.
 }
\label{Fig:entropy}
\end{figure}
%

\subsection{Non-relativistc domain wall during the domination}

Here, we note resultant quantities
 if the domain wall energy density scales as $a^{-1}$ during the domination.

\subsubsection{Matter-dominated era to domain wall dominated era}


At the domain wall decay time $\Gamma_{DW}^{-1}$,
 the ratio of these energy densities is estimated as 
\begin{eqnarray}
 \left. \frac{\rho_M}{\rho_{DW}}\right|_{\Gamma_{DW}} = \left( \frac{\Gamma_{DW}}{H_{eq}}\right)^4 ,
\end{eqnarray}
 from $H \propto a^{-1/2}$, where we assume $\rho_{DW} \propto a^{-1}$
 during the domain wall domination between $H_{eq}$ and $\Gamma_{DW}$.
After the domain walls decay, the energy density of the matter is diluted as
\begin{eqnarray}
\frac{\rho_{M}}{s} = \frac{ 3 T_d}{4}\left( \frac{\Gamma_{DW}}{H_{eq}}\right)^4
 \simeq \frac{ 3 T_d}{4}\left( \frac{\pi^2 g_*(T_d) T_d^4}{10} \frac{M_P^2}{\sigma^2} \right)^2 ,
\label{eq:rho/s:Scaling}
\end{eqnarray}
 for the case that the domain wall decays earlier than the matter does.

\subsubsection{Radiation-dominated era to domain wall dominated era}

Assuming $\rho_{DW} \propto a^{-1}$ during the domain wall domination,
%
 the entropy density ratio of after- to before-domain wall decay is given by 
\begin{eqnarray}
 \Delta = \frac{s_{after}}{s_{before}} \simeq \frac{T_{eq}}{T_d}\left( \frac{H_{eq}}{\Gamma_{DW}}\right)^4
 \simeq  \left(\frac{10 \sigma^2}{\pi^2 g_*(T_d) T_d^4 M_P^2} \right)^{9/4}
 \left(\frac{g_*(T_d)}{g_*(T_{eq})}\right)^{1/4},
\label{rad:delta1-2:Scaling}
\end{eqnarray}
 for $\Delta \gg 1$.
We can obtain an entropy production 
\begin{eqnarray}
 \Delta \simeq 600 \left( \frac{\sigma^{1/3}}{ 50 \,{\rm TeV}}\right)^{27/2}
\left( \frac{2\,{\rm MeV}}{T_d}\right)^9 .
\label{rad:delta2:Scaling}
\end{eqnarray}

\section{Cosmological implications}

In this section, we study implications of the NMSSM domain wall decay to some relics in several models.


\subsection{Thermal relic WIMP LSP such as singlino or sneutrino}

WIMPs have been regarded as a promising dark matter candidate in our Universe.
In the NMSSM, neutralino is the candidate~\cite{Ellwanger:2009dp}. 
In a right-handed neutrino extended model, right-handed sneutrino also becomes
 a WIMP dark matter candidate~\cite{Cerdeno:2008ep}.
Since the WIMP thermal relic abundance is inversely proportional to
 its thermal averaged annihilation cross section $\langle\sigma v\rangle$ as
\begin{equation}
\Omega_{WIMP}h^2 \simeq \frac{0.1\, {\rm pb}}{\langle \sigma v \rangle},
\end{equation}
 too small annihilation cross section leads to overabundant WIMPs.
The Singlino- or Bino-like neutralino,
 or right-handed sneutrino with small couplings is indeed such a case.
The domain wall decay produces extra entropy with
 the dilution factor (\ref{rad:delta1-2:Scaling2}) and
 could regulate the WIMP relic abundance to be
\begin{equation}
\Omega_{WIMP}h^2 \frac{1}{\Delta} \simeq 0.1 ,
\end{equation}
 even for a small annihilation cross section $\langle \sigma v \rangle \ll 1$ pb.

\subsection{The moduli problem in the mirage mediation scenario}

Mirage mediation models appear free from
 the cosmological moduli problem because a moduli mass is quite large.
However, nonthermally produced LSP through a decay chain by way of gravitino
 are in fact overabundant.
Let us examine whether the dmain wall decay dilute those LSPs.

Moduli decay before the energy density of domain walls dominates the Universe,
 because the moduli decay rate
\begin{equation}
\Gamma_{moduli} \simeq \frac{m_{moduli}^3}{8\pi M_P^2},
\end{equation}
 is larger than $H_{\rm eq}$ given by Eq.~(\ref{eq:H-eq:Scaling:MD})
 in the mirage mediation scenario.
At $H\simeq \Gamma_{moduli}$,
 the moduli decay at a moduli dominated Universe produces gravitinos as
\begin{eqnarray}
Y_{3/2} = \frac{n_{3/2}}{s } = B_{3/2}\frac{3T_D}{2m_{moduli}} ,
\end{eqnarray}
 with the branching ratio of moduli decay into gravitinos
$B_{3/2} = {\cal O}(0.01)-{\cal O}(1)$ \cite{Endo:2006zj}, 
 and the Universe becomes radiation dominated.
Here $T_D$ is the decay temperature of the moduli field given by
\begin{eqnarray}
 3 M_P^2 \Gamma_{moduli}^2 = \frac{\pi^2 g_*(T_D)}{30}T_D^4 .
\label{def:TD}
\end{eqnarray}

The entropy density ratio of after- to before-domain wall decay is given
 by Eq.~(\ref{rad:delta1-2:Scaling2}).
Unstable gravitinos decay into LSP with $n_{3/2}= n_{LSP}$ due to R-parity conservation.
Usually, this leads to the overproduction of LSP whose abundance exceeds
 the dark matter abundance. 
After extra entropy production by the domain wall decay,
 the resultant final LSP abundance becomes
\begin{eqnarray}
\frac{\rho_{LSP}}{s} \simeq \frac{3 m_{LSP} T_D}{2m_{moduli}}\frac{B_{3/2}}{\Delta} ,
\end{eqnarray}
 in other words,
\begin{eqnarray}
\Omega_{LSP}h^2 \simeq 4.2 \times 10^8
  \frac{m_{LSP} T_D}{m_{moduli}}\frac{B_{3/2}}{\Delta} \,{\rm GeV}{}^{-1} .
\label{def:delta}
\end{eqnarray}
In figure 4, we consider the case that the LSP is the dark matter, and plot 
$\Omega_{LSP}h^2 = 0.1$ 
 by using (\ref{def:delta}).
 The input parameters are $\lambda=\kappa=0.01$, $T_d=3~{\rm MeV}$, $m_{{\rm LSP}}=100~{\rm GeV}$, 
$m_{{\rm moduli}}=1000~{\rm TeV}$.\\
%
\begin{figure}[ht]
\begin{center}
\epsfig{figure=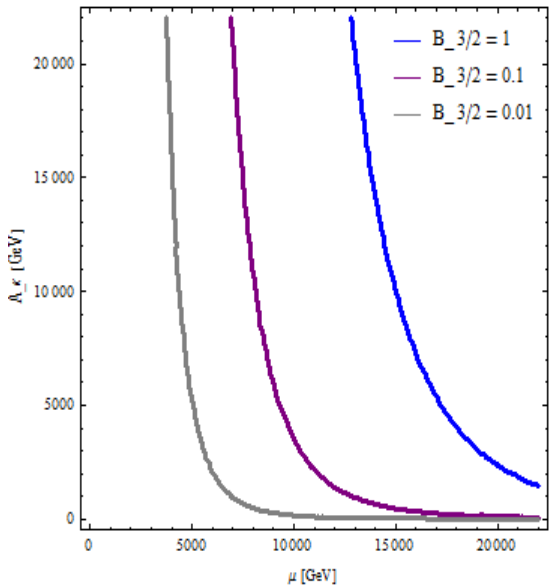, width=8cm,height=8cm,angle=0}
\end{center}
\caption{The required branching ratio contour to keep $\Omega_{LSP}h^2 = 0.1$ 
 in the mirage mediation scenario 
for $\lambda=\kappa=0.01$, $T_d=3~{\rm MeV}$, $m_{{\rm LSP}}=100~{\rm GeV}$, $m_{{\rm moduli}}=1000~{\rm TeV}$.
Above each curve, the relic abundance is smaller than $\Omega_{LSP}h^2 = 0.1$. }
\label{Fig:LSP}
\end{figure}
%

\subsection{The decay constant of the QCD axion}

Finally, we comment on the QCD axion, $a$, with the decay constant $f_a$.
After the QCD transition,
 axions are produced by coherent oscillation, so-called misalignment mechanism, and
 a good candidate for dark matter because its lifetime is much longer than the age of the Universe.
Its abundance is proportional to $f_a^{7/6}$~\cite{Sikivie:2006ni}.
The condition $\Omega_{a} \lesssim \Omega_{DM}$ is rewritten as
\begin{eqnarray}
f_a \lesssim 10^{12} \, {\rm GeV}.
\label{bound:fa}
\end{eqnarray}
$f_a$, which is larger than (\ref{bound:fa}), corresponds to the overproduction of axions.
Again, the domain wall decay can dilute the axion abundance for such a larger $f_a$~\cite{Kawasaki:2004rx}.

For example, with the dilution (\ref{rad:delta1-2:Scaling2}) by the domain wall decay,
 the bound on $f_a$ is relaxed as
\begin{eqnarray}
f_a \lesssim 10^{16} \,{\rm GeV} ,
\end{eqnarray}
 for $\sigma^{1/3}=300$ TeV and $T_d=2$ MeV.

The GUT scale axion decay constant is allowed, which is remarkable.
In superstring theory, the natural decay constant of axionic parts in a closed string moduli 
would be of the order of the GUT scale or string scale~\cite{Svrcek:2006yi}~\footnote{Even larger decay constants 
can be obtained in a certain situation (see e.g., Ref.~\cite{Abe:2014pwa}).}.
Such stringy axions with larger decay constant can be the QCD axion.



\section{Cosmological implications for $w=-1/3$ domain walls}

In this section, we study implications of the NMSSM domain wall decay with  $w=-1/3$.
for  the moduli problem within the gravity mediation scenario.

Now, let us study the dilution of moduli to avoid the moduli problem.
After inflation,
 the moduli would start to oscillate and dominate the energy density of the Universe.
They may decay during or after the BBN and chage the success of BBN.
To avoid such a situation, the energy density of moduli must satisfy
\begin{eqnarray}
\frac{\rho_{moduli}}{s} \lesssim c \cdot 3.6 \times 10^{-9}\, {\rm GeV},
\label{eq:moduli-constraint}
\end{eqnarray}
 where $c \sim 10^{-2} - 10^{-4}$ for 10 TeV moduli mass depending on 
 the coupling between the moduli and the gauge field \cite{Asaka:1999xd}.
We use $c=10^{-3}$ in the following analysis.

The decay of domain walls can dilute the moduli density, which is given as 
\begin{eqnarray}
\frac{\rho_{moduli}}{s}
 \simeq \frac{ 3 T_d}{4}\left( \frac{\pi^2 g_*(T_d) T_d^4}{10} \frac{M_P^2}{\sigma^2} \right)^2,
\label{rho/s:Scaling}
\end{eqnarray}
 as derived in Eq.~(\ref{eq:rho/s:Scaling}). 
It depends on only $T_d$ and tension $\sigma$, which depends on $\lambda$, $\kappa$, $A_\kappa$ and $\mu$.
Imposing the constraint (\ref{eq:moduli-constraint}) on the resultant abundance (\ref{rho/s:Scaling}), we find
\begin{eqnarray}
 \sigma^{1/3} \gtrsim 220 \,{\rm TeV}\left(\frac{10^{-3}}{c}\right)^{1/12} \left(\frac{T_d}{3\, {\rm MeV}}\right)^{3/4} ,
\end{eqnarray}
 where $g(T_d)=10$ is used.

Figure 5 shows the constraints (\ref{eq:moduli-constraint}) with (\ref{rho/s:Scaling}) for $\lambda=\kappa=0.01$, $T_d=3~{\rm MeV}$. The shaded region is excluded by the constraint.\\
%
\begin{figure}[ht]
\begin{center}
\epsfig{figure=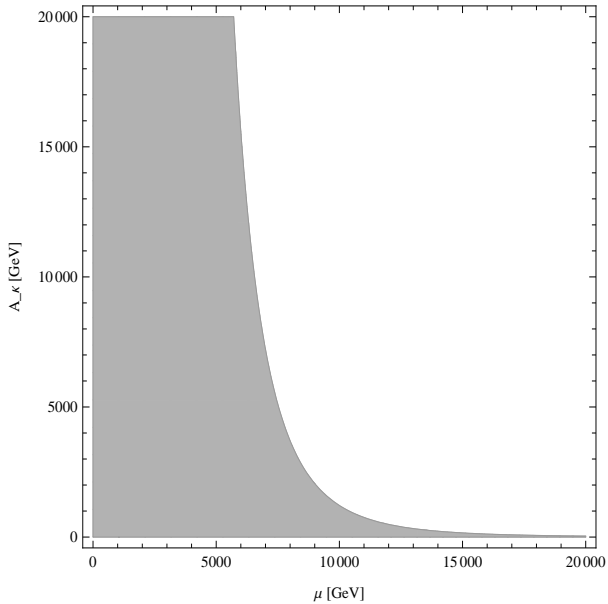, width=8cm,height=8cm,angle=0}
\end{center}
\caption{The bound of the moduli abundance in gravity mediation scenario 
for $\lambda=\kappa=0.01$, $T_d=3~{\rm MeV}$. 
The yellow region is allowed in the $(\mu, A_{\kappa})$ plane. 
 }
\label{Fig:moduli}
\end{figure}
%

\section{Conclusion and discussion}

We have studied the cosmological implication of unstable domain walls
 in the NMSSM.
The spontaneous breaking of the $Z_3$ discrete symmetry in the NMSSM
 causes the cosmological domain wall problem.
We consider that the $Z_3$ symmetry is slightly but explicitly broken and 
the domain walls decay with the decay temperature $T_d$.
The domain walls easily dominate the density of the Universe 
and its decay causes a late-time entropy production, 
depending on its tension $\sigma$ and $T_d$.
Such entropy production has significant implications in thermal history.
They can dilute unwanted relics such moduli, gravitino, LSP and axion.

We have shown that $T_d$ of several MeV
 dilute various relics in several scenarios.
Those includes thermal WIMP LSP in gravity mediation model,
 nonthermally produced LSP in mirage mediation and
 misalignment produced cold axion in Peccei-Quinn extented models.
If the energy density of domain wall network decreases as $\rho_{DW}\propto a^{-1}$
 during domain wall domination,
 cosmological moduli problem in gravity mediation also might be relaxed.
%
%


\section*{Acknowledgments}
This work was supported in part by the Grant-in-Aid for Scientific Research 
No.~25400252 and No.~26247042 (T.K.) and on Innovative Areas No.~26105514 (O.S.)
 from the Ministry of Education, Culture, Sports, Science and Technology in Japan. 
%




\end{document}